# Launching of Visible-Range Hyperbolic Polaritons by Gold Nanoantennas in a natural van der Waals crystal

Clara Clemente-Marcuello[1, 2], Haozhe Tong[1,3], Kirill V. Voronin[1], Pablo Alonso-González[2,4,†], Alexey Y. Nikitin[1, 5,†]

[1] Donostia International Physics Center (DIPC), Donostia–San Sebastián 20018, Spain.
[2] Department of Physics, University of Oviedo, Oviedo 33006, Spain.
[3] Universidad del País Vasco/Euskal Herriko Unibertsitatea (UPV/EHU), 48940, Leioa, Spain.
[4] Center of Research on Nanomaterials and Nanotechnology, CINN (CSIC–Universidad de Oviedo), El Entrego 33940, Spain.
[5] IKERBASQUE, Basque Foundation for Science, Bilbao 48013, Spain.

[†] Corresponding authors. Email: **pabloalonso@uniovi.es**; **alexey@dipc.org**

## ABSTRACT

Anisotropic van der Waals materials provide a powerful platform for nanoscale optoelectronics, enabling strong light–matter interaction and deep electromagnetic field confinement mediated by polaritons, hybrid light–matter excitations with unique dispersion properties. While polaritonic phenomena in van der Waals heterostructures have been extensively explored in the mid-infrared frequency range, their behaviour at the visible frequencies remains largely unexplored, in part due to the lack of knowledge on natural materials supporting anisotropic and highly confined visible-range polaritons. In this context, $MoOCl_2$, an anisotropic van der Waals metal, is particularly interesting, since it supports hyperbolic plasmon polaritons (PPs) that enable directional propagation and subwavelength light compression. Here, we investigate the strategy for launching anisotropic PPs in $MoOCl_2$ in the visible frequency range using gold rod nanoantennas. The nanoantennas, placed on top of the $MoOCl_2$ crystal, excite in-plane anisotropic PP modes, effectively overcoming the momentum mismatch between waves in free-space and nanoscale PPs. We demonstrate a strong electromagnetic field confinement, angle-dependent absorption, and controlled anisotropic PP launching enabled by gold nanoantennas, highlighting the potential of $MoOCl_2$ as a compact platform for nanoscale waveguiding and optical signal processing. By providing a practical antenna-based strategy for exciting visible-range PPs, this work addresses the lack of compact elements for optical signal manipulation and opens new opportunities for optoelectronic devices based on van der Waals polaritonics.

## INTRODUCTION

Recent advances in van der Waals (vdW) materials have revealed a rich landscape of anisotropic polaritons, thus providing a novel approach for light manipulation at the nanoscale[1–3]. These hybrid light-matter excitations emerge from the strong coupling between electromagnetic fields and collective dipolar oscillations in layered crystals, resulting in an unprecedented level of field confinement. In particular, the emergence of hyperbolic and strongly anisotropic polaritonics modes allows for the manipulation of light at extreme subwavelength scales, enabling fundamental optical phenomena such as ray-like propagation[4–6], diffractionless canalization[7–12], anomalous reflection and refraction,[13–15], or deeply subwavelength focusing[16–18], which are unattainable in conventional isotropic media. To date, these anisotropic polaritons have been extensively mapped across a broad spectral range, with the majority of landmark effects demonstrated primarily at mid-infrared (mid-IR) and terahertz (THz) frequencies[19]. In this spectral regime, several vdW materials have emerged as natural platforms supporting hyperbolic polaritons, including α-$MoO_3$[20], $V_2O_5$[21], and $WTe_2$[22]. In particular, α-$MoO_3$ has been extensively studied as a prototypical natural hyperbolic crystal, where strong in-plane dielectric anisotropy gives rise to hyperbolic isofrequency contours and highly anisotropic polariton dispersion [5,14,20].

The translation of the above-mentioned anisotropic polaritonic phenomena to the visible frequency range remains a significant challenge. Up to now, the achievement of visible-range hyperbolic response has been primarily accomplished through the utilisation of artificial metamaterials, such as metal-dielectric multilayers or nanowire arrays[23]. However, these platforms require complex and sophisticated nanofabrication techniques, and the incorporation of nanostructured metals introduces substantial ohmic losses, which limit polariton propagation[24]. Consequently, there is a clear need for natural van der Waals crystals capable of supporting highly confined, anisotropic polaritons in the visible range without the drawbacks of artificial nanostructuring[1,3]. In this context, van der Waals metal $MoOCl_2$ has recently emerged as a singular natural platform supporting hyperbolic response in the visible frequency range[25–28]. This material is of particular interest as it is characterised by strong electronic correlations and structural asymmetry, leading to a significant intrinsic optical anisotropy[26,27]. The existence of such properties is known to enable the existence of visible-frequency hyperbolic or generally anisotropic plasmon polaritons (PPs), which offer deep subwavelength confinement and crystallographically defined directional propagation [5,14]. By supporting PPs in the visible regime, $MoOCl_2$ provides a natural solution to the challenges of nanostructuring and high losses associated with traditional artificial metamaterials previously mentioned and offers additional opportunities due to its potential compatibility with conventional optoelectronic circuits.

Despite the remarkable properties of visible-range hyperbolic PPs in $MoOCl_2$, their extreme subwavelength confinement results in a significant momentum mismatch with free-space light, hindering direct excitation. To overcome this, we employ optical metallic rod nanoantennas to bridge the momentum gap, a strategy successfully pioneered for launching polaritons in graphene[29], h-BN[30–32], and α-$MoO_3$[33] at infrared frequencies.

In this work, we demonstrate that this antenna-based strategy can be effectively transferred to the visible frequency range. Analyzing the dielectric permittivity and hyperbolic PPs dispersion relations of $MoOCl_2$, we systematically investigate the PPs launching in thin $MoOCl_2$ layers by gold rod nanoantennas. In particular, we study angle-dependent absorption and the launching efficiency as a function of the antenna parameters. Our results establish a robust framework for the controlled excitation of highly confined hyperbolic PPs in a natural van der Waals material at visible frequencies.

**RESULTS**:

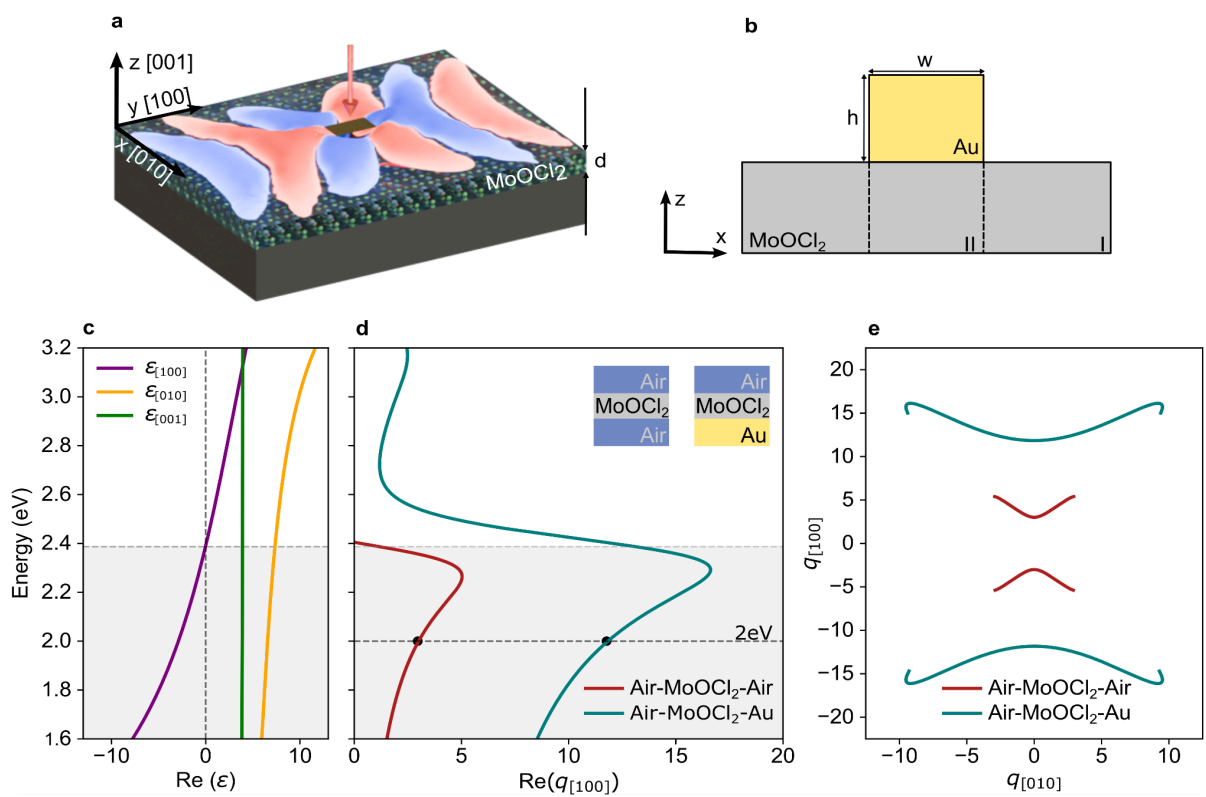

**Figure 1: Dispersion of PP in a MoOCl₂ layer** (a) Schematics of the sample, the gold rod antenna on top of the MoOCl₂ layer placed on the SiO₂ substrate, illuminated by a plane wave (red arrow). Under the incident radiation, the antenna excites hyperbolic PPs in the layer. (b) Schematics of the view in the $zx$-plane indicating setup I (Air-MoOCl₂-Air) and II (Au-MoOCl₂-Air) for a continuous layer. (c) Real part of the dielectric permittivity tensor as a function of frequency for MoOCl₂. A colored grey background indicates the range where in-plane hyperbolic PPs are supported. (d) Dispersion relation for PPs propagating in setups I and II along the [100] crystal axis. (e) Isofrequency curves for PPs in these structures for a selected photon energy of 2 eV. In all panels, the layer thickness is 20 nm.

The main concept of the studied structure is illustrated by schematics in Fig. 1a. We consider a MoOCl₂ layer of thickness $d$ = 20 nm placed on a dielectric substrate and coupled to a gold nanorod antenna. The 3D color plot represents the full-wave numerical simulations of the spatial distribution of the vertical component of the electric field, generated under illumination by a plane wave incident from the top half-space. As shown in Fig. 1b (presenting a cross-section of the geometry illustrated in Fig. 1a), in the presence of the Au rod, our geometry can be divided into two sub-regions: free-standing MoOCl₂ (I) and MoOCl₂ covered by gold (II). Notably, if the Au antenna thickness is larger than the skin depth, the Au layer in region (II) can be approximated by a semi-infinite Au half-space. Fig. 1b also defines the coordinate system and clarifies that panels (d) and (e) correspond to calculations performed for a homogeneous in-plane MoOCl₂ layer rather than a strip of finite width.

To understand and explain the generated electric field pattern, we first proceed to visualize the dispersion of anisotropic PPs in a MoOCl₂ layer. Taking into account the large momentum of PP, their dispersion relation can be calculated from the following approximate explicit expression for the normalized in-plane momentum [5,34]:

$$q = \frac{\rho}{k_0 d}\left(\arctan\frac{\varepsilon_1 \rho}{\varepsilon_z} + \arctan\frac{\varepsilon_3 \rho}{\varepsilon_z} + \pi l\right), \quad l = 0,1,2\ldots$$

Here, $k_0$ is the wavevector of light in free space; $\varepsilon_1$ and $\varepsilon_3$ are the dielectric permittivities of the substrate and supersubstrate; $d$ is the thickness of the MoOCl₂ layer; and $l$ denotes the mode index. The parameter $\rho$ is given by:

$$\rho = i\sqrt{\frac{\varepsilon_z q^2}{\varepsilon_x q_x^2 + \varepsilon_y q_y^2}} = i\sqrt{\frac{\varepsilon_z}{\varepsilon_x \cos^2\alpha + \varepsilon_y \sin^2\alpha}},$$

where $\alpha$ is the polar angle between the direction of propagation of PP and the x-axis, and $\varepsilon_i$ are the components of the anisotropic dielectric permittivity tensor of MoOCl₂ with $i = x, y, z$. The dielectric permittivity of Au is taken from [35], while the permittivity tensor of MoOCl₂ is described using the model reported in [25–28].

Fig. 1c shows the real part of the dielectric permittivity tensor of MoOCl₂[27,28]. The crystallographic directions [100] and [010] correspond to the $y$- and $x$-axes, respectively. To interpret the behavior of PPs in this composite system, we first analyze the dispersion properties of PPs propagating in two in-plane homogeneous three-layer structures: (I) Air–MoOCl₂–Air and (II) Air–MoOCl₂–Au, according to the above-mentioned splitting of our geometry.

Fig. 1d displays the dispersion relations of PPs propagating in both structures as a function of the in-plane momentum $q_{[100]}$. These dispersion relations reveal high-$k$ PP modes with momenta far exceeding that of free-space photons at the same energy, evidencing strong electromagnetic confinement in the visible regime. Based on this dispersion analysis, we then select specific frequencies of interest at which the PP response is further analyzed. Fig. 1e shows the isofrequency curves (IFCs) at a fixed energy of 2 eV for both configurations, calculated using Eq. (1). The IFCs exhibit a pronounced hyperbolic shape, indicating strongly direction-dependent polariton propagation. Compared to the Air–MoOCl₂–Air configuration, the presence of Au significantly modifies both the shape and extent of the IFCs, pushing the PP modes to much larger in-plane wavevectors.

These features originate from the optical anisotropy of MoOCl₂, leading to PP modes with direction-dependent dispersion[14]. When a MoOCl₂ layer is covered with gold, the dispersion of PPs is strongly modified due to the electric field screening by the Au substrate, which results in PP modes with larger in-plane momentum and higher vertical field confinement[36,37]. The high dispersion contrast between the polaritons in the free-standing MoOCl₂ and the gold-covered MoOCl₂ ensures an efficient coupling between free-space light and the hyperbolic PP modes, opening access to visible-range PPs in MoOCl₂[32]. These dispersion properties underlie the directional propagation and excitation phenomena investigated in this work.

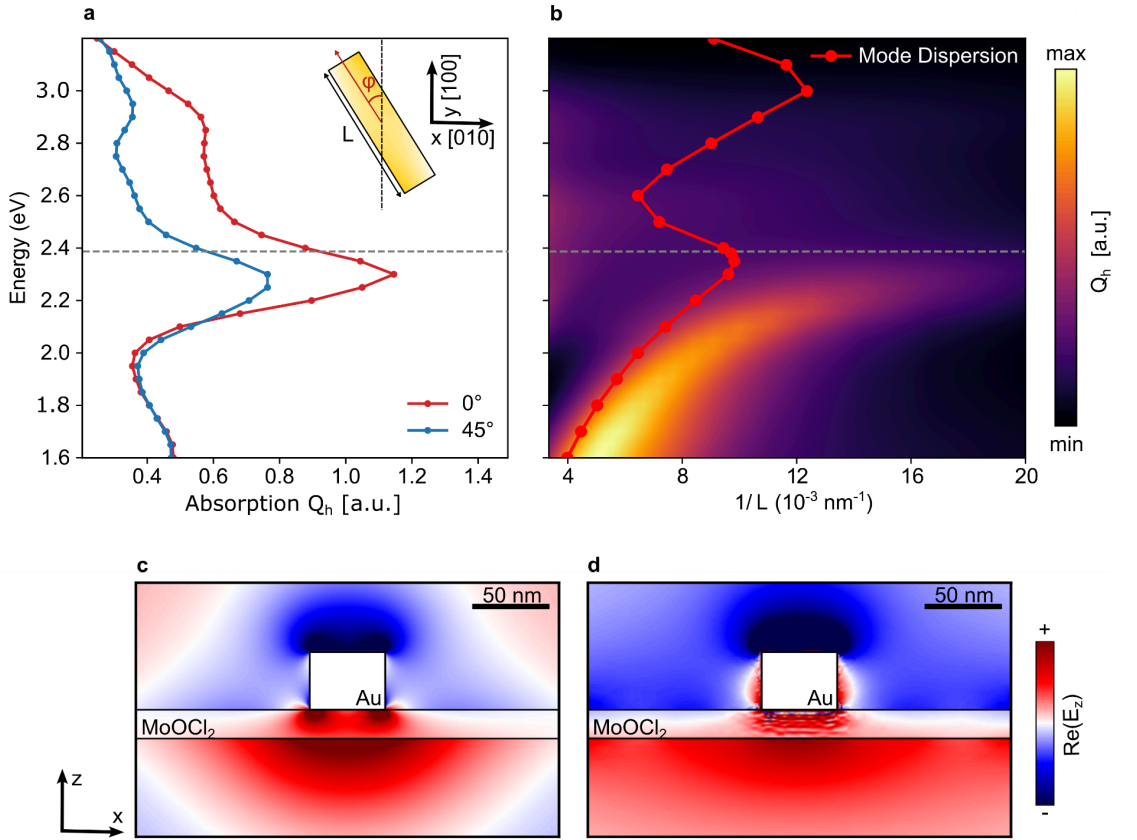

**Figure 2: Gold rod antennas on MoOCl$_2$ slab for different angles and antenna lengths.** (a) Simulated absorption by the antenna of the length $L$ = 70 nm for two different orientation angles as a function of the incident light frequency. The inset shows a top-view schematic of the antenna placed on the MoOCl$_2$ layer, where $\varphi$ denotes the antenna orientation angle with respect to the crystallographic axes. (b) Colorplot of the total absorption in the system of the gold rod antenna on MoOCl$_2$ as a function of the antenna length and the incident frequency obtained by the full-wave numerical simulation. The red curve represents the dispersion of the mode propagating along the antenna, shown in panel c. The dashed lines mark the transition between hyperbolic (1.6-2.4 eV) and elliptic (2.4-3.2 eV) dispersion regimes. (c) Spatial distribution of the vertical component of the electric field of the PP mode propagating along the infinitely long antenna at the frequency of 1.6 eV. (d) Distribution of the vertical component of the electric field in the cross-section of the real antenna, taken at the end of the antenna, including the full antenna height and width, obtained by full three-dimensional simulation, illustrating the excitations of the mode shown in c.

To study the excitation efficiency of PP modes by rod nanoantennas, we analyze the antenna-assisted absorption of the incident light. For this, we perform a full-wave numerical simulation of the system, consisting of a 40-nm-high and 50-nm-wide gold rod nanoantennas of different lengths, $L$, placed on top of a 20-nm-thick MoOCl$_2$ layer, illuminated by a normally incident plane wave. The antenna orientation is varied within the plane of the layer, forming an angle $\varphi$ with the crystallographic [100]-axis of MoOCl$_2$. The insert in Fig. 2a shows a top-view schematic defining the antenna geometry, its orientation angle, and the crystallographic axes used in the simulations. The total antenna-assisted absorbed power is calculated as the absorption inside the gold antenna added to the absorption by the MoOCl$_2$ layer in the presence of the antenna and subtracted by the absorption of the bare $MoOCl_2$ layer, in an analogous structure without the antenna. This definition isolates the contribution associated with antenna-PP coupling, similarly to [32].

Fig. 2a shows the frequency-dependent absorption for two antenna orientations, corresponding to angles of 0° and 45° with respect to the [100]-axis. The absorption exhibits a peak at the frequency of PP resonance in $MoOCl_2$, especially pronounced when the antenna is aligned along the [100]-axis. The latter case corresponds to the direction in which the real part of the permittivity is negative in the 1.6–2.4 eV range. The angular dependence of the absorption reflects the anisotropic response of $MoOCl_2$, where an efficient coupling occurs only when the antenna is aligned along directions that support propagating polaritonic modes. To better understand the excitation of the PPs, we plot a color map of the total absorption as a function of photon energy and the inverse antenna length, $1/L$, in Fig. 2b. The horizontal gray dashed line marks the transition between the hyperbolic (1.6–2.4 eV) and elliptic (2.4–3.2 eV) regimes of in-plane PP propagation. The absorption features are strongest within the hyperbolic regime and are suppressed upon entering the elliptic regime.

To interpret the observed absorption maxima, we develop a reduced two-dimensional (2D) model to compute the dispersion of the PP mode propagating along the antenna, assuming one in-plane dimension, parallel to the y-axis, to be infinite. Fig. 2c displays a two-dimensional vertical electric field distribution, $E_z$, at 1.6 eV, illustrating the waveguiding mode propagating along the gold rod placed on top of the $MoOCl_2$ layer. We further assume that the waveguiding PP mode experiences multiple reflections from the antenna ends ("bouncing" between the antenna's edges) so that a Fabry-Perot (FP) resonance of the PP waveguiding mode is built, similarly to [30–32]. The condition of the FP resonance reads as $Lk_w + \psi = n\pi$, with $k_w$ is the in-plane wavevector of the PP waveguiding mode, $\psi$ is the reflection phase, and $n$ is an integer. The obtained polaritonic dispersion, namely, $k_w/\pi$ (assuming $n = 1$ and $\psi = 0$), is shown by the red curve, and superimposed on the absorption map in Fig. 2b. The absorption maximum follows a clear dispersion trend as a function of the inverse antenna length, $1/L$, and energy, proving the role of the FP resonance of the antenna waveguide mode in the formation of the absorption maximum. The deviation between the absorption maxima and the dispersion of a modeled PP waveguiding structure is observed in the lower-energy region. This mismatch can be attributed to the non-zero value of the frequency-dependent phase, $\psi$ [32]. The field distributions shown in Fig. 2c and Fig. 2d further confirm the excitation of a PP waveguiding mode confined within the $MoOCl_2$ layer. Panel (c) corresponds to the reduced two-dimensional model, which assumes translational invariance along the antenna and describes mode propagation along its axis. Panel (d) shows the corresponding field profile obtained from full three-dimensional simulations. Both panels exhibit the same characteristic field distribution, demonstrating that the field in the realistic antenna model can be represented as the PP waveguide mode field[19].

The strong correlation between absorption, antenna geometry, and polaritonic dispersion establishes absorption as a reliable probe of polariton launching. These results highlight the importance of anisotropy and antenna orientation in achieving controlled excitation of visible-range PPs, laying the groundwork for the efficient launching and directional propagation analyzed in the following.

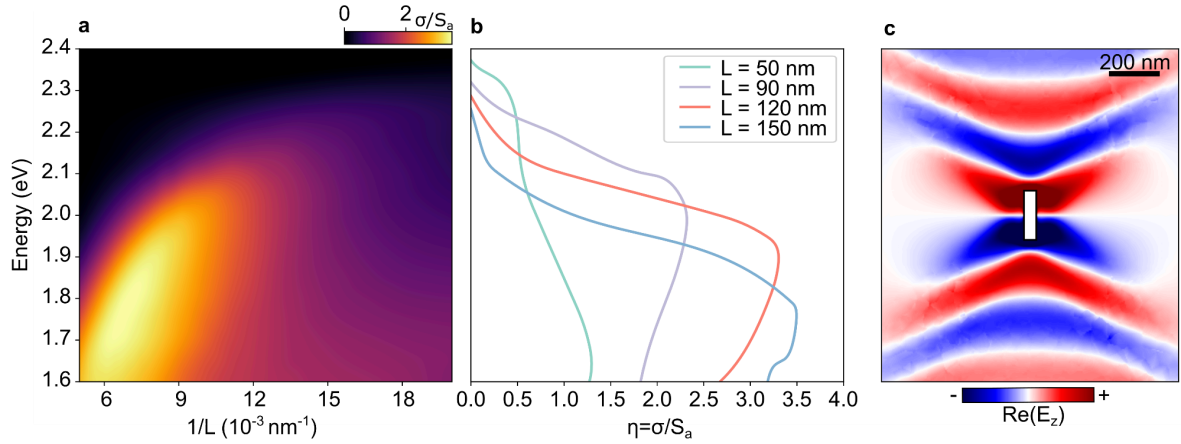

**Figure 3: Launching efficiency of PPs for different gold antenna lengths.** (a) Launching efficiency of PPs normalized to the top-view cross-section area of the antenna, $\eta = \sigma/S_a$, as a function of the inverse antenna length, $1/L$, and incident photon energy. (b) Normalized launching efficiency, $\eta$, for selected antenna lengths ($L$ = 50, 90, 120 and 150 nm) as a function of energy. (c) Distribution of the vertical component of the electric field, $E_z$, for the PP mode launched by a gold nanoantenna. The simulation is performed at an excitation energy of 1.6 eV for an antenna length of 200 nm, showcasing the characteristic directional propagation of the hyperbolic modes in the $xy$-plane.

To further quantify the efficiency of PP launching mediated by gold nanoantennas on $MoOCl_2$, we analyze the absorbed power associated exclusively with the PPs in $MoOCl_2$ emanating away from the antenna. To that end, we calculate the launching efficiency by extracting the absorption associated with launched plasmon polaritons (PPs) from the total absorption, excluding local electromagnetic losses. Namely, the absorption is calculated by excluding the volume of $MoOCl_2$ located directly beneath the antenna and subtracting the background absorption of the layer without the antenna. The resulting absorbed power is normalized by the intensity of the incident light and the geometrical cross-sectional area of the antenna, resulting in the launching efficiency defined as $\eta = P_{abs}/I_{inc} S_a$ [32]. Using full-wave numerical simulations, we evaluate the launching efficiency as a function of inverse antenna length, $1/L$, and frequency.

Fig. 3a displays a color map of the normalized launching efficiency within the hyperbolic regime, while Fig. 3b presents line profiles for selected antenna lengths (50, 90, 120 and 150 nm) for a better visibility of the efficiency spectral features. Consistently with our previous analysis, the data reveal a pronounced resonant behavior, with the launching efficiency exhibiting a maximum at specific combinations of antenna length and incident frequency. At these resonance conditions, the efficiency increases by nearly a factor of four compared to off-resonant configurations.

When the antenna length is tuned to match the FP resonance condition, it efficiently converts incident radiation into propagating PPs, overcoming momentum mismatch. This process is directly visualized in Fig. 3c, which shows the out-of-plane electric field distribution, $E_z$, at 1.6 eV for a 200 nm antenna. The spatial profile reveals a directional polaritonic response propagating away from the antenna, with characteristic "V-shaped" wavefronts consistent with the hyperbolic dispersion of MoOCl$_2$ at this energy. These results demonstrate that antenna geometry provides a practical and tunable way to optimize PP launching, confirming the effectiveness of gold nanoantennas as couplers for visible-range polaritonics of anisotropic van der Waals materials.

**CONCLUSIONS**

In this work, we have investigated the excitation and launching of visible-range PPs in the anisotropic van der Waals material MoOCl$_2$. By analyzing its dielectric permittivity and dispersion properties and performing full-wave numerical simulations of the angle-dependent absorption, we have demonstrated that gold rod nanoantennas provide an effective and compact strategy for coupling free-space radiation into high-momentum polaritonic modes. The antenna geometry and orientation are crucial for controlling excitation conditions, while resonant antenna lengths enable a four-fold enhancement of the polariton launching efficiency. By isolating the contribution of launched PPs from local absorption, we have quantitatively established the conversion of incident radiation into propagating polaritonic modes via the nanoantenna.

These results highlight the critical role of crystal anisotropy and antenna design in achieving controlled polariton excitation in vdW materials. The antenna-based launching approach demonstrated here addresses a key limitation in visible-range polaritonics and positions MoOCl$_2$ as a promising platform for nanoscale waveguiding and optical signal processing. More broadly, this work opens new opportunities for integrated optoelectronic and nanophotonic devices operating at visible frequencies by exploiting the advantages of hyperbolic materials.

**Funding Sources**

The study was funded by the Department of Science, Universities and Innovation of the Basque Government (grant PIBA-2023-1-0007) and the IKUR Strategy; by the Spanish Ministry of Science and Innovation (grant PID2023-147676NB-I00 and PID2022-141304NB-I00). P. A.-G. acknowledges support from the European Research Council under Consolidator (grant no. 101044461).